# *Emulating photonic time interfaces via smooth temporal transitions*

*Mariya Antyufeyeva[1,2], and Victor Pacheco-Peña[1*]*

[1] *School of Mathematics, Statistics and Physics, Newcastle University, Newcastle Upon Tyne, NE1 7RU, United Kingdom*
[2] *V. N. Karazin Kharkiv National University, 4, Svobody Square, Kharkiv, 61022, Ukraine*

**The introduction of time as an additional degree of freedom to control wave-matter interactions have opened new avenues to fully control wave propagation in four dimensions ($x,y,z,t$). Time interfaces (rapid changes of the constitutive relations of the medium where a wave propagates) have recently become popular as they are the temporal analogue of spatial interfaces. While recent groundbreaking experimental work has demonstrated time interfaces from water waves, microwaves and the optical regime, rapidly changing, for instance, the permittivity of the medium requires carefully engineered structures. Here, we study the possibility of implementing smooth temporal transitions of the refractive index of the medium in order to mimic the response of time interfaces via adiabatic modulations. It is shown that indeed, as long as the signal is inside the structure during the modulation time, there are some values of rising/falling time of the adiabatic modulation that enables a full approximate emulation of a time interface. These results may open further avenues to explore by relaxing the speed at which the time-modulation should be introduced when designing four-dimensional media.**

---

[*] email: <victor.pacheco-pena@newcastle.ac.uk>

## Introduction

Controlling wave-matter interactions in three-dimensional space has been at the centre of the microwaves and photonics community, making it possible to manipulate wave propagation using spatially engineered structures placed along the path where a wave propagates[1–12]. Recently, wave propagation in four dimensions, with time $t$ as the extra degree of freedom, has become a hot research topic, enabling the manipulation of waves in both space and time[13–21]. Wave propagation in temporally modulated media (in particular rapid changes using a single step function) was theoretically explored by Morgenthaler in a seminal work[3] where the electromagnetic properties of the medium were changed rapidly from one value to a different value at a time $t_0$ using a step function, i.e., a time-dependent relative permittivity $\varepsilon(t)$ and/or relative permeability $\mu(t)$. Such scenario is now known as a temporal boundary/interface where, similar to a spatial interface between two materials of different constitutive parameters (such as $\varepsilon$, $\mu$, impedance), a time reflected (backward, BW, wave) and a time refracted (forward, FW, wave) are excited. These two waves will then travel with a frequency different than the signal existing inside the medium for times $t < t_0$[3,22].

The exploration of time interfaces has led to exciting new physical phenomena[22–26], including temporal aiming[27], temporal multilayer structures modelling higher-order transfer functions[28] and impedance transformers that can be used as temporal filters[29], classical and quantum antireflection temporal coating[30,31], temporal analogue of metasurface unit cells[32], broadband light sculpting[33], analogue of Faraday polarization rotation at time interface in magnetoplasma[34], broadband frequency translation[35], the phase-conjugation of time-reflected waves[36], temporal prism[37] and inverse prism[38,39], temporal Brewster angle[40], temporal twistronics[41], amplification and holding of surface waves[42–44], ultra slow surface waves[44], propagating to evanescent waves coupling[45], among others[18,46–53]. Interestingly, the first experimental demonstrations of time interfaces have been reported in the past few years. Examples include time interfaces for water waves[54], microwaves[35,36] and the optical regime[55–57], demonstrating how the field is rapidly evolving towards experimental implementations.

One of the challenges of time interfaces is that the electromagnetic properties of the medium where the wave propagates should rapidly change with a fall/rise time smaller than the period of the signals already present in the medium[52]. However, recent works have shown that tapered time interfaces (using smooth profiles[58] or multisteps[59]) can also be implemented for achieve, for instance, frequency conversion where the final frequency can be similar as that obtained when using a sharp time interface (as long as the final permittivity value is the same, as expected). Such an approach can be viewed as the spatial tapering used in horn antennas where reflections due to an abrupt spatial change between the guided structure and free-space are minimized[60]. Similarly, temporal tapering produces a FW wave with the BW being reduced. An interesting question is, what will happen to the phase of the FW wave depending on the rise/fall time of the smoothly modulated refractive index $n(t)$? i.e., is there a value of rise/fall time of a smoothly changing time-dependent material where the phase profile of the FW wave is exactly (or at least approximately) the same as the phase of the FW wave produced by a step function of $n(t)$? If such conditions exist, it would be possible to mimic a time interface fully using smoothly varying material properties, potentially opening further practical implementations.

Inspired by the interesting opportunities and possibilities that time interfaces can open, in this work we address this question by theoretically investigating the conditions under which the smooth profile of $n(t)$ can mimic the amplitude and phase values for the electromagnetic wave produced by a time interface (single step function of $n(t)$). To do this, we make use of the Modal Basis Method, also known as the Evolutionary Approach to Electromagnetics (EAE) in the time domain [61–65] considering the time domain electromagnetic field of a cavity resonator surrounded by perfect electric conducting walls and filled with a time-varying medium. The refractive index, $n(t)$, of the medium filling the cavity is changed in time both using a step function and a sigmoid function in order to modify the rise/fall time of the modulated material. For simplicity, we consider values of $n(t)$ such that the impedance is preserved for all times. As it will be shown below, this approximation enables us to map the instantaneous frequency, amplitude and phase of the FW wave, or total field in our case. The analytical solutions obtained for the time part of the field (modal amplitudes) allow us to analyse in detail its reliance on the time duration of the adiabatic modulation and contrast of the medium (variation of $n$ before and after the temporal modulation). Interestingly, it will be shown that there are values of rise/fall times of $n(t)$ where the steady-state field oscillations of the FW wave (after the time modulation has been completely applied) have approximately the same amplitude and phase as that of the FW wave produced by a single step function of $n(t)$. Full analytical expressions of these conditions are provided, demonstrating how, in applications such as frequency conversion, it is possible to closely mimic a time interface using sigmoid/smooth functions of $n(t)$. The theoretical calculations are confirmed using numerical simulations via COMSOL Multiphysics®.

## Results

**Description of the problem under study**

The schematic representation of the problem under study is shown in Fig. 1. We consider a cavity resonator (surrounded by a perfect electric conductor, PEC, walls) filled with an isotropic, lossless, homogeneous medium having a time-dependent refractive index $n(t)$ (Fig. 1a). A time interface is implemented by rapidly changing the constitutive parameters of the medium where a wave travels ($n(t)$ in our case). This is schematically shown in the inset of Fig. 1a and Fig. 1c, where the time-dependent $n(t)$ of the medium filling the cavity has a value of $n_1$ for times $t < t_0$ and it is rapidly changed to $n_2$ at a time $t = t_0$ (i.e., a single step function of $n$). In this manuscript, it is considered that the impedance of the medium is kept constant for all times ($Z = \sqrt{\mu_1/\varepsilon_1} = \sqrt{\mu_2/\varepsilon_2}$, with $\varepsilon_{1,2}$, $\mu_{1,2}$ as the relative permittivity and permeability, respectively, of the medium before and after time interface is introduced). Here, in addition to time interfaces, we will study smooth transitions of $n(t)$ from $n_1$ to $n_2$ (Fig. 1d) in order to theoretically establish similarities between rapid (time interface) and adiabatic temporal modulations. Importantly, the cavity resonator from Fig. 1 has, in general, an arbitrary shape and is geometrically defined by a volume $V$ which is bounded by the closed PEC surface $S$ (Fig. 1a).

For completeness, the schematic representation of the refractive index of the cavity at different time instants is shown in Fig. 1c,d for the case of a single step function of $n(t)$ and a smooth transition (with a transition time $\Delta t$), respectively. The impact of introducing such temporal functions of $n(t)$ is schematically shown in Fig. 1b, where it can be observed how the amplitude of the field already inside the cavity may change, but the spatial distribution (for example, the fundamental mode of the cavity) is preserved. As it will be discussed below, the single step or smooth temporal modulation of $n(t)$ will influence the amplitude and phase of the electromagnetic field in the cavity. Mathematically, throughout this manuscript, the single step function and smooth temporal transition of the refractive index of the medium are defined using the constitutive $\varepsilon$ and $\mu$ values, as follows, respectively:

$$\varepsilon(t)^{\text{step}} = \varepsilon_1 + (\varepsilon_2 - \varepsilon_1)H(t-t_0) = \begin{cases} \varepsilon_1, & t < t_0 \\ \varepsilon_2, & t > t_0 \end{cases}, \quad (1a)$$

$$\varepsilon(t)^{\text{smooth}} = \varepsilon_1 + \frac{\varepsilon_2 - \varepsilon_1}{1 + e^{-\gamma(t-t_0)}}, \quad (1b)$$

where $\varepsilon(t)^{\text{step}}$ and $\varepsilon(t)^{\text{smooth}}$ are the time-dependent relative permittivity for the time interface (single step function) and smooth transition, respectively, $\mu(t) = Z^2 \varepsilon(t)$, $n(t) = \sqrt{\mu(t)\varepsilon(t)} = Z\varepsilon(t)$, $H(t-t_0)$ is the Heaviside function centred at $t_0$ (corresponding to the time when the transition is half-way from $\varepsilon_1$ to $\varepsilon_2$ for the smooth function), $\Delta t$ is the rise/fall time of the smooth transition (Eq. 1b) defined as $\Delta t = \zeta/\gamma$, with $\gamma$ as control parameter (when $\gamma$ goes to infinity (Eq. 1b) converges into (Eq. 1a)) and $\zeta$ as the parameter used here to estimate the transition time. As detailed below, this parameter ensures that the smooth transition has approximately the same values of $\varepsilon(t)$ as the time interface (at times $t < t_0$ and $t > t_0$ from Eq. 1a).

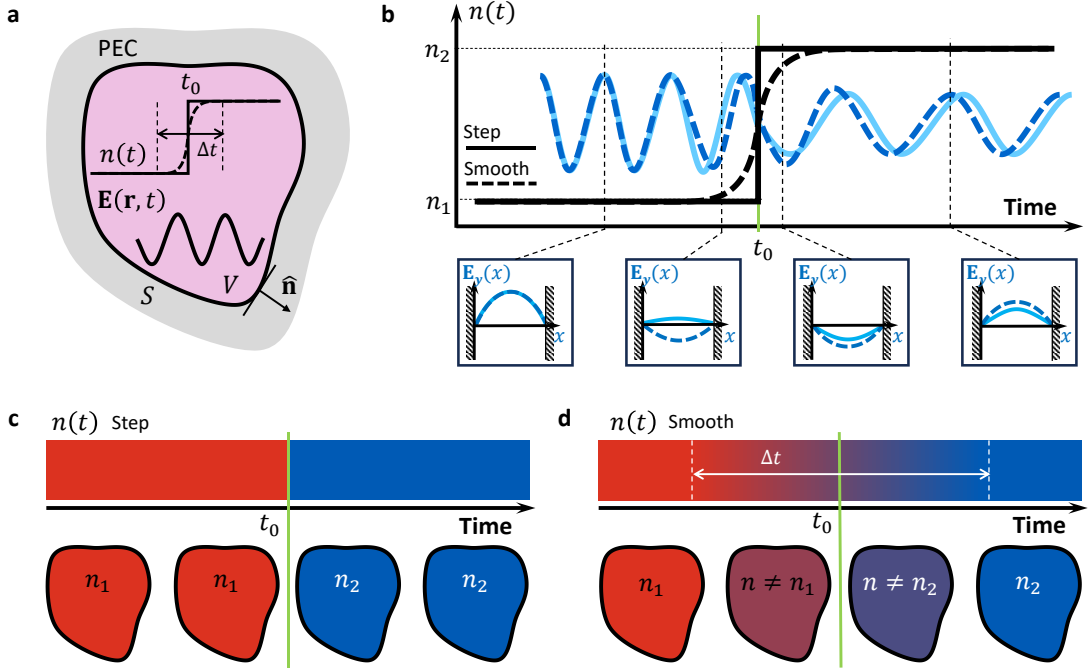

**Fig. 1 Schematic representation of a time interface (single step) and smooth change of refractive index of the medium inside a cavity resonator.** Schematic of the general cavity resonator geometry and problem statement (**a**), representation of the problem in time for the step and smooth change of refractive index and its influence on the field inside the cavity (**b**). Solid light blue and dashed blue lines represent the field distribution for the case of a time interface and a smooth function of $n(t)$, respectively. (**c-d**) Sketches of the time interface and corresponding smooth change of $n(t)$.

**Theory**

In this section, we present the theoretical analysis of the scenario shown in Fig. 1 using EAE in the time domain. It is important to note that the theory behind expanding electromagnetic fields within cavities considering certain normal modes with orthogonality properties was introduced and has been subject of study since the last century [61,63–72]. In the present section, we provide the key steps and results of the solution for the specific electromagnetic problem under study.

As it is known, when considering time interfaces, the constitutive relations describing the interaction between the electromagnetic field and the medium inside the cavity (considering nondispersive materials where the working frequency falls away from any material resonances [3,22,26,73]) can be written as $\mathbf{D}(\mathbf{r},t) = \varepsilon_0 \varepsilon(t) \mathbf{E}(\mathbf{r},t)$, $\mathbf{B}(\mathbf{r},t) = \mu_0 \mu(t) \mathbf{H}(\mathbf{r},t)$ where $\mathbf{D}(\mathbf{r},t)$, $\mathbf{B}(\mathbf{r},t)$, $\mathbf{E}(\mathbf{r},t)$, $\mathbf{H}(\mathbf{r},t)$ are the electric field displacement, magnetic flux density, electric field and magnetic field, respectively, $\mathbf{r}$ is the position vector, $t$ is time and $\varepsilon_0$, $\mu_0$ are the absolute permittivity and permeability values in free space. In this work we consider that the medium inside the cavity is lossless. Moreover, it is considered that for times $t < 0$ there is an electromagnetic signal that exists within the cavity. This source is then stopped at $t = 0$. This means that there is an initial distribution of the electromagnetic field $\mathbf{E}(\mathbf{r},0) = \mathbf{E}_0(\mathbf{r})$, $\mathbf{H}(\mathbf{r},0) = \mathbf{H}_0(\mathbf{r})$ at the time when the fields are started to be observed ($t = 0$). For times $t > 0$ the electromagnetic signal keeps resonating within the cavity and at a time $t = t_0 > 0$ a time interface is introduced. Note that this configuration is in line with [13,18]. For simplicity we specify the shape of the cavity as rectangular (see Fig. 1b), but we note that this technique can be applied to any geometry of the cavity. Following the EAE in the time domain, in general, we are looking for the electric and magnetic field in the cavity filled with the time-dependent medium as the expansion in the modal basis vectors [65]

$$\mathbf{E}(\mathbf{r},t) = \sum_{m=1}^{\infty} e'_m(t) \mathbf{E}'_m(\mathbf{r}) + \sum_{m=1}^{\infty} e''_m(t) \mathbf{E}''_m(\mathbf{r}) + \sum_{\alpha=1}^{\infty} a_\alpha(t) \nabla \phi_\alpha(\mathbf{r}), \qquad (2a)$$

$$\mathbf{H}(\mathbf{r},t) = \sum_{m=1}^{\infty} h'_m(t) \mathbf{H}'_m(\mathbf{r}) + \sum_{m=1}^{\infty} h''_m(t) \mathbf{H}''_m(\mathbf{r}) + \sum_{\beta=1}^{\infty} b_\beta(t) \nabla \psi_\beta(\mathbf{r}). \qquad (2b)$$

where $\mathbf{E}'_m(\mathbf{r})$, $\mathbf{E}''_m(\mathbf{r})$ are the electric and $\mathbf{H}'_m(\mathbf{r})$, $\mathbf{H}''_m(\mathbf{r})$ are the magnetic solenoidal modal vectors depending only on spatial coordinates and representing the divergence-free electromagnetic part of the field in the cavity. The terms $e'_m(t)$, $e''_m(t)$, $h'_m(t)$, $h''_m(t)$ are the electric and magnetic modal amplitudes representing their evolution with time. The prime (′) and double-prime (″) superscripts denote the different sets of modal vectors. The terms $\nabla \phi_\alpha(\mathbf{r})$ and $\nabla \psi_\beta(\mathbf{r})$ are the electric and magnetic irrotational modal vectors, also depending only on spatial coordinates. These modal vectors account for the presence of static electromagnetic fields in the cavity[74], and $a_\alpha(t)$, $b_\beta(t)$ are their electric and magnetic modal amplitudes. To find the modal vectors, one can consider the well-known eigenvalue problems and scalar Helmholtz equations with Dirichlet and Neumann boundary conditions, taking into account the orthonormality condition[65] as follows:

$$\begin{cases} \nabla \times \mathbf{E}'_m(\mathbf{r}) = \omega'_m \mu_0 \mathbf{H}'_m(\mathbf{r}) \\ \nabla \times \mathbf{H}'_m(\mathbf{r}) = \omega'_m \varepsilon_0 \mathbf{E}'_m(\mathbf{r}) \\ (\hat{\mathbf{n}} \times \mathbf{E}'_m(\mathbf{r}))|_S = 0 \end{cases} \quad (3a)$$

$$\begin{cases} \nabla \times \mathbf{H}''_m(\mathbf{r}) = \omega''_m \varepsilon_0 \mathbf{E}''_m(\mathbf{r}) \\ \nabla \times \mathbf{E}''_m(\mathbf{r}) = \omega''_m \mu_0 \mathbf{H}''_m(\mathbf{r}) \\ (\hat{\mathbf{n}} \cdot \mathbf{H}''_m(\mathbf{r}))|_S = 0 \end{cases} \quad (3b)$$

$$\begin{cases} (\nabla^2 + \kappa_\alpha^2)\phi_\alpha(\mathbf{r}) = 0, \quad \phi_\alpha(\mathbf{r})|_S = 0 \\ \frac{\kappa_\alpha^2 \varepsilon_0}{V} \int_V |\phi_\alpha(\mathbf{r})|^2 dV = 1 \end{cases} \begin{cases} (\nabla^2 + \varkappa_\beta^2)\psi_\beta(\mathbf{r}) = 0, \quad \left(\frac{\partial}{\partial \mathbf{N}}\psi_\beta(\mathbf{r})\right)\Big|_S = 0 \\ \frac{\varkappa_\beta^2 \mu_0}{V} \int_V |\psi_\beta(\mathbf{r})|^2 dV = 1 \end{cases} \quad (3c)$$

$$\frac{\varepsilon_0}{V} \int_V |\mathbf{E}'_m|^2 dV = \frac{\mu_0}{V} \int_V |\mathbf{H}'_m|^2 dV = \frac{\varepsilon_0}{V} \int_V |\mathbf{E}''_m|^2 dV = \frac{\mu_0}{V} \int_V |\mathbf{H}''_m|^2 dV = 1,$$
$$\frac{\varepsilon_0}{V} \int_V |\nabla \phi_\alpha|^2 dV = \frac{\mu_0}{V} \int_V |\nabla \psi_\beta|^2 dV = 1 \quad (3d)$$

where $\omega'_m \neq 0$, and $\omega''_m \neq 0$, $\kappa_\alpha > 0$, $\varkappa_\beta > 0$, are real eigenvalues. The eigenvalues $\omega'_m$ and $\omega''_m$ represent the eigenfrequencies of an empty resonator (i.e., where the filling material is vacuum). When the shape of the cavity is rectangular, $\omega'_m = \omega''_m \equiv \omega_{pqs} = \frac{\pi}{\sqrt{\varepsilon_0 \mu_0}} \sqrt{\left(\frac{p}{l_x}\right)^2 + \left(\frac{q}{l_y}\right)^2 + \left(\frac{s}{l_z}\right)^2}$, where $p, q, s$ are integers indicating the number of variations in the standing wave pattern in the $x$, $y$, $z$ directions, respectively, and $l_x, l_y, l_z$ are the dimensions of the cavity along each direction[60]. The solutions to Eq. 3a,b are associated with complete sets of solenoidal TE and TM modes[61,65,74].

Coming back to the description of the initial field in the considered cavity ($\mathbf{E}_0(\mathbf{r})$ and $\mathbf{H}_0(\mathbf{r})$), we can reduce our calculation by considering that only one TE mode exists within the cavity and no static field is present. This means that the irrotational electric and magnetic parts of the initial fields associated with the presence of a static field are zero[75] (i.e., in last terms from Eq. 2, $a_\alpha(0) = 0$ and $b_\beta(0) = 0$). Due to this, the expansions from Eq. 2 for the initial fields are limited only by the first sum and only by the first term in this sum (i.e., $m = 1$). As it is known, as the magnetic field is $\pi/2$ out-of-phase with the electric field in a lossless resonator, the magnetic field becomes zero twice per period of field oscillation while, at the same time, the electric field reaches its maximum within entire volume of the cavity. It is this moment of time that we call $t = 0$ in this work and, as explained above, is the time when the source stops. Based on this, we can define $\mathbf{H}_0(\mathbf{r}) = 0$. With these considerations, the expressions of the electric and magnetic field using the expansion in the modal basis vectors (Eq. 2) are reduced to:

$$\mathbf{E}(\mathbf{r}, 0) = \mathbf{E}_0(\mathbf{r}) = e'_1(0)\mathbf{E}'_1(\mathbf{r}) = e_0 \mathbf{E}'_1(\mathbf{r})$$
$$\mathbf{H}(\mathbf{r}, 0) = \mathbf{H}_0(\mathbf{r}) = 0,$$
(4)

with $t = 0$, $e_0 = \frac{\varepsilon_0}{V}\int_V \mathbf{E}_0(\mathbf{r}) \cdot \mathbf{E}'_1(\mathbf{r})dV$. The term $e_0 \neq 0$ plays the role of the initial condition of the solenoidal electric modal amplitude $e'_m(t)$ [61] when $m = 1$. Eq. 4 then represents the modal expansion of initial field at any point $\mathbf{r}$ inside the cavity. Naturally, the initial conditions for all modal amplitudes except of $e'_1(t)$ are zero as we are dealing only with the fundamental TE mode, as mentioned above.

Regarding the modal amplitudes $e'_m(t), e''_m(t), a_\alpha(t), h'_m(t), h''_m(t), b_\beta(t)$ for times $t > 0$ from Eq. 2, as they describe the time evolution of the electric and magnetic field at each point in space within the volume of the cavity, they can be obtained by projecting Maxwell's equations onto the modal basis vectors. A detailed description of similar procedure was presented in [61,76]. The application of this procedure for current problem yields the following equations for finding the modal amplitudes (also called evolutionary equations in the literature[61,63–65,67,74,77]):

$$\begin{cases} \frac{d}{dt}\varepsilon(t)e'_m(t) - \omega'_m h'_m(t) = 0, & e'_1(0) = e_0, \quad e'_{m \neq 1}(0) = 0, \quad m = 1, 2, \ldots \\ \frac{d}{dt}\mu(t)h'_m(t) + \omega'_m e'_m(t) = 0, & h'_m(0) = 0 \quad m = 1, 2, \ldots \end{cases}$$
(5a)

$$\begin{cases} \frac{d}{dt}\varepsilon(t)e''_m(t) - \omega''_m h''_n(t) = 0, & e''_m(0) = 0 \quad m = 1, 2, \ldots \\ \frac{d}{dt}\mu(t)h''_m(t) + \omega''_m e''_m(t) = 0, & h''_m(0) = 0 \quad m = 1, 2, \ldots \end{cases}$$
(5b)

$$\frac{d}{dt}\varepsilon(t)a_\alpha(t) = 0, \quad a_\alpha(0) = 0 \quad \alpha = 1, 2, \ldots$$
(5c)

$$\frac{d}{dt}\mu(t)b_\beta(t) = 0, \quad b_\beta(0) = 0 \quad \beta = 1, 2, \ldots$$
(5d)

The solution of the homogeneous linear differential equation with constant coefficients accompanied with zero initial conditions yields zero sought functions[78]. Thus, over the sets of evolutionary equations from Eq. 5, only Eq. 5a for $m = 1$ has a nonzero solution. This fact simplifies the electric and magnetic field expansion from Eq. 2 in the considered cavity in a similar fashion to the initial field decomposition from Eq. 4. Based on this, the electric and magnetic field for the scenario presented in Fig. 1 are determined as:

$$\mathbf{E}(\mathbf{r},t) = e'_1(t)\mathbf{E}'_1(\mathbf{r}), \quad \mathbf{H}(\mathbf{r},t) = h'_1(t)\mathbf{H}'_1(\mathbf{r}), \tag{6}$$

where, $\mathbf{E}'_1(\mathbf{r})$ and $\mathbf{H}'_1(\mathbf{r})$ are the electric and magnetic field distribution of the fundamental TE mode in the cavity, determined by Eq. 3a, and $e'_1(t)$, $h'_1(t)$ are the temporal amplitudes of the electric and magnetic field described by the system of evolutionary equations:

$$\begin{cases} \frac{d}{dt}\varepsilon(t)e'_1(t) - \omega'_1 h'_1(t) = 0, & e'_1(0) = e_0, \\ \frac{d}{dt}\mu(t)h'_1(t) + \omega'_1 e'_1(t) = 0, & h'_1(0) = 0. \end{cases} \tag{7}$$

The effective approach to solve systems like Eq. 7 can be found in [61,63,74], and using those steps, the solution of the system from Eq. 7 is obtained as follows:

$$e'_1(t) = \frac{\varepsilon(0)}{\varepsilon(t)} e_0 \cos[\omega'_1 \tau(t)], \tag{8a}$$

$$h'_1(t) = -\frac{\varepsilon(0)}{Z\varepsilon(t)} e_0 \sin[\omega'_1 \tau(t)], \tag{8b}$$

where $\tan^{-1}\varphi_0 = \frac{Zh_0}{e_0}$, $\mu(t) = Z^2\varepsilon(t)$ and $\tau(t) = \int_0^t \frac{du}{Z\varepsilon(u)}$ is an auxiliary time variable. Thus, Eq. 8 represent a full analytical expression of the temporal behaviour of the electromagnetic fields at any point $\mathbf{r}$ of the cavity. Now, let us denote the time-varying amplitude in Eq. 8 as $\frac{\varepsilon(0)}{\varepsilon(t)} = A(t)$ and the phase as $\varphi(t) = \omega'_1 \tau(t)$. Importantly, this follows the definition of the instantaneous frequency $[\Omega(t)]$ as defined in [3]; i.e., $\Omega(t) = \frac{d\varphi(t)}{dt} = \omega'_1 \frac{d\tau(t)}{dt} = \omega'_1 \frac{1}{Z\varepsilon(t)} = \frac{\omega'_1}{n(t)}$.

Since the goal of solving the problem is to estimate the contribution of the smooth transition of $n(t)$ (Eq. 1b) compared to a time interface (Eq. 1a), the parameters $A(t)$, $\tau(t)$ and instantaneous frequency $\Omega(t)$ for the "step" and "smooth" changes of $n(t)$ for times $t < t_0$ (before the time interface) and $t > t_0$ (after the time interface) can be calculated as:

$$A(t)^{\text{step}} = \begin{cases} 1, & t < t_0 \\ \varepsilon_1/\varepsilon_2, & t > t_0 \end{cases} \quad \Omega(t)^{\text{step}} = \begin{cases} \frac{\omega'_1}{Z\varepsilon_1}, & t < t_0 \\ \frac{\omega'_1}{Z\varepsilon_2}, & t > t_0 \end{cases}$$

$$\tau(t)^{\text{step}} = \begin{cases} \frac{t}{\varepsilon_1 Z}, & t < t_0 \\ \frac{1}{\varepsilon_2 Z}\left\{t + \frac{\varepsilon_2 - \varepsilon_1}{\varepsilon_1} t_0\right\}, & t > t_0 \end{cases} \tag{9a}$$

$$A(t)^{\text{smooth}} = \begin{cases} \frac{(\varepsilon_2+\varepsilon_1 e^{\gamma t_0})(1+e^{\gamma(t_0-t)})}{(1+e^{\gamma t_0})(\varepsilon_2+\varepsilon_1 e^{\gamma(t_0-t)})}, & t < t_0 \\ \frac{(\varepsilon_2+\varepsilon_1 e^{\gamma t_0})(1+e^{-\gamma(t-t_0)})}{(1+e^{\gamma t_0})(\varepsilon_2+\varepsilon_1 e^{-\gamma(t-t_0)})}, & t > t_0 \end{cases} \qquad \Omega(t)^{\text{smooth}} = \begin{cases} \frac{\omega_1'}{Z}\frac{1+e^{-\gamma(t_0-t)}}{\varepsilon_1+\varepsilon_2 e^{-\gamma(t_0-t)}}, & t < t_0 \\ \frac{\omega_1'}{Z}\frac{1+e^{-\gamma(t-t_0)}}{\varepsilon_2+\varepsilon_1 e^{-\gamma(t-t_0)}}, & t > t_0 \end{cases}$$

(9b)

$$\tau(t)^{\text{smooth}} = \begin{cases} \frac{1}{\varepsilon_1 Z}\left\{t - \frac{\varepsilon_2-\varepsilon_1}{\varepsilon_2}\left[\frac{1}{\gamma}\ln\left(\frac{\varepsilon_2 e^{-\gamma(t_0-t)}+\varepsilon_1}{\varepsilon_2 e^{-\gamma t_0}+\varepsilon_1}\right)\right]\right\}, & t < t_0 \\ \frac{1}{\varepsilon_2 Z}\left\{t + \frac{\varepsilon_2-\varepsilon_1}{\varepsilon_1}t_0 - \frac{\varepsilon_2-\varepsilon_1}{\varepsilon_1}\left[\frac{1}{\gamma}\ln\left(\frac{\varepsilon_2+\varepsilon_1 e^{-\gamma(t-t_0)}}{\varepsilon_2 e^{-\gamma t_0}+\varepsilon_1}\right)\right]\right\}, & t > t_0 \end{cases}$$

As observed, the parameters from Eq. 9a correspond to those obtained by Morgenthaler[3] and widely known in the literature, considering the case when the impedance of the medium is constant. Regarding the terms describing the smooth transition (Eq. 9b), it can be seen how for longer times ($t \gg t_0$) the final frequency of the electromagnetic signal inside the cavity using a smooth transition of refractive index is approximately the same as that of a step function, as expected. In the next section we provide a detailed theoretical and numerical analysis of both smooth and step temporal transitions of $n(t)$, of the material filling the cavity. Moreover, we present an in-depth analysis of the conditions under which the full expressions from Eq. 9a,b for the time interface and smooth transition, respectively, may approximately coincide.

**Comparison between time interfaces and smooth temporal transitions**

Once we have obtained the full analytical expression of the modal amplitudes within the cavity, we can now proceed to compare the solutions when using a step function and a smooth transition of $n(t)$. The aim here is to answer the question: what will happen to the phase of the FW wave depending on the rise/fall time of the smoothly modulated refractive index $n(t)$? As described in Eq. 9, the instantaneous frequency $\Omega(t)$ for a smooth transition of $n(t)$ becomes approximately the same as that of the step function for longer times after the temporal modulation has been introduced. However, the question about the phase remains unanswered. This section provides some physical insights to answer this question in order to investigate the possibility to emulate time interfaces via a smooth transition of $n(t)$. Here, the theoretical calculations using smooth and step functions of $n(t)$ (Eqs. 8-9) are presented showing both the amplitude and phase characteristics of the FW wave in the time domain (noting that, as mentioned above, we keep impedance unchanged through the process, so the FW wave is indeed the total field inside the cavity). These results are then compared with numerical simulations via COMSOL Multiphysics®, for completeness[79].

As mentioned in the previous section, the capability of the EAE in the time domain to completely split the spatiotemporal problem into spatial (Eq. 3) and temporal (see Eqs. 5,7) problems allows us to obtain Eqs. 8-9 which fully describe the temporary behaviour of the electromagnetic field within the cavity. Importantly, only the first eigenvalue $\omega'_1$ ($\omega'_m$ with $m = 1$ in the boundary eigenvalue problem from Eq. 3a) connects the evolutionary equations from Eq. 5a and, consequently, from Eq. 7, with the spatial problem from Eq. 3. Due to this, without loss of generality, we can consider a one-dimensional (1D) cavity with vertical parallel plate metallic walls (PEC in our case) as the one shown in Fig. 1b by considering the corresponding 1D value of $\omega'_1$ in Eq. 8-9. i.e., the corresponding eigenvalue is $\omega'_1 = \pi\sqrt{\left(\frac{1}{l_x}\right)^2/(\mu_0\varepsilon_0)}$ where $l_x$ is the distance between plates.

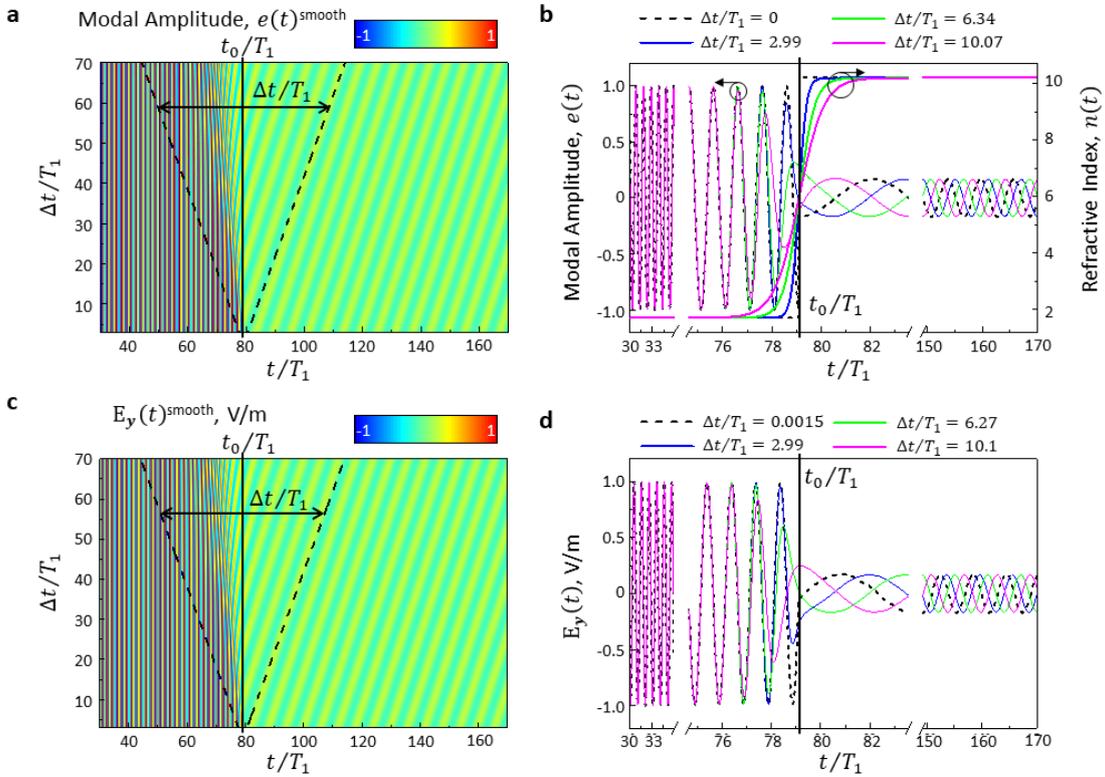

**Fig. 2 Time-dependent fields inside the cavity using smooth transitions and time interfaces via $n(t)$: from smaller to larger values.** Theoretical results of the electric modal amplitude (**a**) and numerical results of the electric field (**c**) as a function of rising time and normalized time using smooth transitions of $n(t)$ from $\varepsilon_1 = 2$ to $\varepsilon_2 = 12$ when the impedance is time-independent, $Z = 0.85$. (**b,d**) theoretical and numerical results, respectively, extracted from (**a,c**) using different values of rising time $\Delta t/T_1$ (solid lines) via a smooth transition of $n(t)$ along with the results of a single time interface (step function, dotted line). For the numerical simulations, the cavity has a length $l_x = 117.57$mm corresponding to half-wavelength inside the material filling the cavity at times before the time interface is applied. The vertical solid line in all panels correspond to the value when the time interface is applied. The diagonal dashed lines from (**a,c**) delimit the duration of the transitions of $n(t)$ when implementing a smooth transition.

With this in mind, let us first study the case when $n(t)$ is changed from a smaller to a larger value (all values larger than 1). The time interface is introduced at $t_0/T_1$. For the smooth transitions we consider different rising times $\Delta t/T_1$ with $T_1 = 2\pi n_1/\omega'_1 = 2\pi Z\varepsilon_1/\omega'_1$ as the period of the field oscillation at times before $n(t)$ is changed (with instantaneous frequency of electromagnetic field at these times defined as $\omega'_1/Z\varepsilon_1$). Specifically, for this study we use $T_1 = 0.67$ns (corresponding to a frequency of ≈1.5 GHz) with $\Delta t = \zeta/\gamma$, ($\zeta = 20$, which is defined to ensure that the exponential in the denominator of Eq. 1b is small, i.e., $e^{-\zeta/2} \ll 1$ at times $t = t_0 \pm \Delta t/2$; see below for a detailed explanation of these times), $t_0 = 53$ ns and $\varepsilon_1 = 2$ and $\varepsilon_2 = 12$, as an example. As mentioned above, we also consider a time dependent $\mu(t)$ with corresponding values of $\mu_1$ and $\mu_2$ chosen accordingly to keep a time-independent impedance of $Z = 0.85$. The numerical simulations were carried out using the time-domain solver from COMSOL Multiphysics® following the same cavity configuration as in[18], further details can be found in the methods section below. Using this setup, the theoretical results (Eqs. 8-9) of the mode amplitude $e(t)^{\text{smooth}}$ as a function of transition time $\Delta t/T_1$ and normalized time $t/T_1$ are shown in Fig. 2a. To guide the eye, the vertical solid line represents the time at which a time interface would be applied ($t_0/T_1 = 79.1$) while the diagonal dashed lines represent the transition time of the smooth function of $n(t)$ ($\Delta t/T_1$). The duration of the smooth transition $\Delta t/T_1$ is changed from 3 to 70 with a step of ≈0.19 (theory) and ≈0.3 (simulations). Importantly, as we keep $\zeta = 20$, the control parameter $\gamma$ is first modified to define different slopes of the smooth transition. $\Delta t$ is then calculated using $\Delta t = \zeta/\gamma$. Effectively, this means that $\Delta t$ will provide information about the rising time of the smooth transition of $n(t)$. This, however, will have a different meaning when $\gamma$ is kept constant and $\zeta$ is changed, as it will be detailed in the next section.

From these results, it can be observed how the frequency of the field inside the cavity is changed once the smooth transition is applied. Moreover, for times after the modulation has finished (results on the right-hand side of the right dashed diagonal line) the frequency seems to be similar for all values of $\Delta t/T_1$, as expected due to $\varepsilon_2$ (and corresponding $\mu_2$) is the same for all the smooth transitions of $n(t)$. To corroborate the validity of these results, we provide the numerical simulations via COMSOL Multiphysics® in Fig. 2c where the results of the time-dependent electric field for the smooth transitions of $n(t)$ $[E_y(t)^{\text{smooth}}]$ are shown. A good qualitatively agreement is observed between both theoretical and numerical results. For completeness, we extracted the values of $e(t)^{\text{smooth}}$ and $E_y(t)^{\text{smooth}}$ for different values of $\Delta t/T_1$ from Fig. 2a,c, respectively, and the results are shown as solid lines in Fig. 2b,d, respectively, along with the value of $n(t)$ for each case (see Fig. 2b). The normalized amplitudes of $e(t)^{\text{smooth}}$ and $E_y(t)^{\text{smooth}}$ are in agreement. The small phase difference between $e(t)^{\text{smooth}}$ and $E_y(t)^{\text{smooth}}$ comes from the fact that, in the numerical simulations, the phase of the signal within the cavity is slightly different to that of the theoretical analysis when the smooth transition is introduced. Such small difference, however, does not affect the analysis as the aim here is to compare the results between the smooth and step functions of $n(t)$ while a comparison between the theoretical and numerical results is just to validate the calculations shown in the previous sections. Finally, and to have the full set of solutions, we provide in Fig. 2b,d the theoretical and numerical results of $e(t)^{step}$ and $E_y(t)^{step}$, respectively, as dotted lines, showing how the phase of the smooth transitions and time interface is different, while the amplitudes are similar. However, one can ask, is there a single value or a set of values of $\Delta t/T_1$ for the smooth transition of $n(t)$ that can provide approximately the same amplitude and phase of the mode amplitude and electric field within the cavity as that of a step function $n(t)$ (time interface)? i.e., can one fully mimic the response of a time interface via adiabatic changes of $n(t)$?

To answer this question, the theoretical values of $e(t)^{step} - e(t)^{smooth}$ and numerical simulation results of $E_y(t)^{step} - E_y(t)^{smooth}$ as a function of $\Delta t/T_1$ and $t/T_1$ are shown in Fig. 3a,d, respectively. Apart from the agreement between the results, it is interesting to notice that there are values of $\Delta t/T_1$ where $e(t)^{step} - e(t)^{smooth} \approx 0$ and $E_y(t)^{step} - E_y(t)^{smooth} \approx 0$. This means that there are indeed values of $\Delta t/T_1$ that may enable the full emulation of a time interface using smooth time-dependent functions of $n(t)$. To calculate the conditions under which both solutions (smooth and step cases) may be the same, we take into account the following: first, it is considered that $t_0 > \Delta t/2$ to ensure the same value of $\varepsilon(t) = \varepsilon_1$ at the beginning of the observation time ($t = 0$) for both the smooth transition of $n(t)$ and the time interface (see some illustrations related to the violation of this condition in the supplementary material where it is shown that the fields never converge as such violation will produce a considerably different value of $\varepsilon(t)$ at $t = 0$ for the smooth transition compared to the value for the time interface). Next, it is also considered that the amplitude and phase matching of the oscillation produced by the smooth transition and the time interface is calculated outside the duration of the transition of $n(t)$ ( i.e. for times not involving the time interval from $t = t_0 - \Delta t/2$ to $t = t_0 + \Delta t/2$ or, in other words, where the permittivity of the medium is approximately either $\varepsilon_1$ or $\varepsilon_2$ but not in between). Importantly, this means that we need to ensure that $\varepsilon(t)$ for the smooth transition at times $t = t_0 - \Delta t/2$ and $t = t_0 + \Delta t/2$ is the same as the time interface (i.e., $\varepsilon_1$ or $\varepsilon_2$, respectively, for the time interface at times $t < t_0$ and $t > t_0$, respectively). To enable this, let us evaluate the first time ($t = t_0 - \Delta t/2$). Using Eq. 1b, $\varepsilon(t)|_{t=t_0-\Delta t/2} =$ $\frac{\varepsilon_2+\varepsilon_1 e^{\gamma\Delta t/2}}{1+e^{\gamma\Delta t/2}} = \frac{\varepsilon_2+\varepsilon_1 e^{\zeta/2}}{1+e^{\zeta/2}}$ meaning that $\varepsilon_2 e^{-\zeta/2} \ll \varepsilon_1$ in order to obtain $\varepsilon(t)|_{t=t_0-\Delta t/2} = \varepsilon_1$ (for the cases shown in this work, as mentioned above, $\zeta = 20$ so $\varepsilon_2 e^{-10} \ll \varepsilon_1$). This means that, with the mathematical definition of $\varepsilon(t)$ for the smooth transition from Eq. 1b, one can simply choose the value of $\zeta$ and then recalculate $\Delta t$ in order to fulfil the condition $\varepsilon(t)|_{t=t_0-\Delta t/2} = \varepsilon_1$. The same process can be applied for the second time, $t = t_0 + \Delta t/2$, where using Eq. 1b $\varepsilon(t)|_{t=t_0+\Delta t/2} = \frac{\varepsilon_2+\varepsilon_1 e^{-\zeta/2}}{1+e^{-\zeta/2}}$ meaning that $\varepsilon_1 e^{-\zeta/2} \ll \varepsilon_2$ in order for the smooth transition to have the same value as the time interface ($\varepsilon(t)|_{t=t_0+\Delta t/2,} = \varepsilon_2$). These two conditions are not mutually exclusive. It is important to note here that as $\zeta$ is finite, the condition $\varepsilon_1 e^{-\zeta/2} \ll \varepsilon_2$ will then mean that $\varepsilon(t)|_{t=t_0+\Delta t/2} \approx \varepsilon_2$. In other words, the parameter $\zeta$ can be used to measure the accuracy of the final permittivity value (and hence the overall solution of the modal amplitude) for the smooth transition compared to the time interface. A discussion about this parameter $\zeta$ will be presented in the following section.

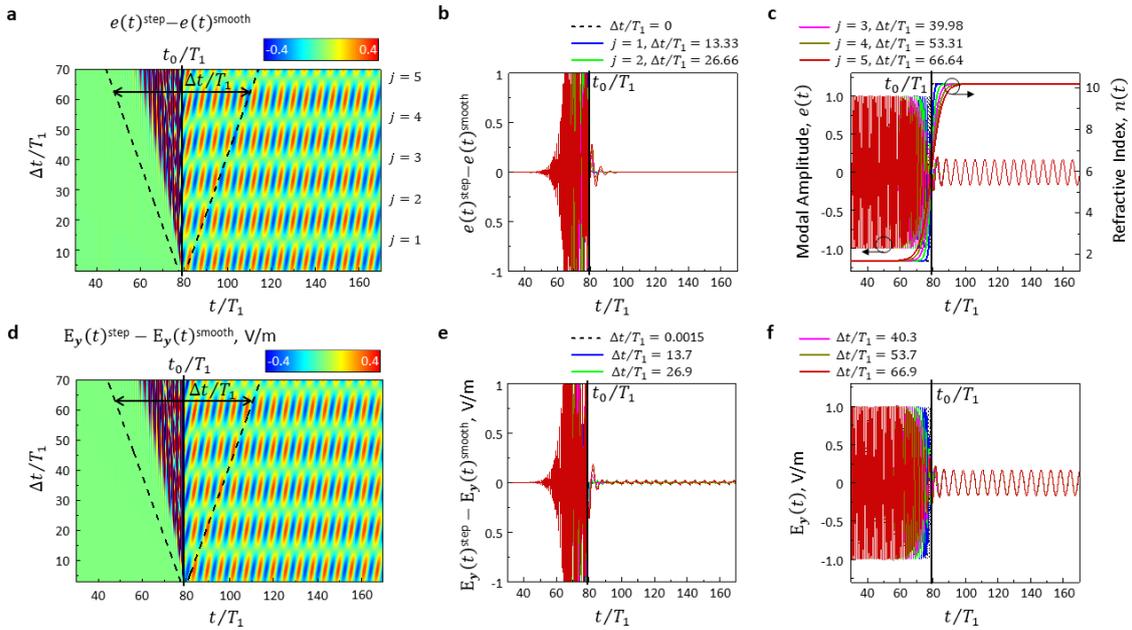

**Fig. 3 Emulating time interfaces via smooth transitions: from smaller to larger values.** Theoretical results of the difference between the electric modal amplitudes produced by step and smooth values of $n(t)$ (**a**) and numerical results of the difference of corresponding electric fields (**d**) as a function of rising time of the smooth transition and time for the scenario described in Fig. 2. (**b,e**) Theoretical and numerical results, respectively, for the values of $\Delta t/T_1$ from (**a,d**) where $e^{step} - e^{smooth} \approx 0$ and $E_y^{step} - E_y^{smooth} \approx 0$, respectively. Theoretical results of electric modal amplitudes modulated by the time interface (step function), dotted line, and smooth transition, solid lines of $n(t)$ (**c**) for the same values of $\Delta t/T_1$ as in panel (**b**). (**f**) Corresponding electric field for the same values of $\Delta t/T_1$ as in panel (**e**). The solid vertical and dashed diagonal lines have the same meaning as those in Fig. 2.

Now, with these considerations, let us analyse the expressions from Eq. 9a,b to understand if the solutions for both smooth transition and time interface are similar. Let us first start with the expression for the amplitude $A(t)$ from Eq. 9 and establish $A(t)^{step} = A(t)^{smooth}$ for longer times ($t > t_0$). The conditions that enable this equality to hold requires $\frac{\varepsilon_1}{\varepsilon_2} \gg e^{-\gamma t_0}$ (see methods). Similar to the analysis for $\varepsilon(t)$ from Eq. 1a,b from the previous paragraph, the relation $\frac{\varepsilon_1}{\varepsilon_2} \gg e^{-\gamma t_0}$ that satisfies $A(t)^{step} = A(t)^{smooth}$ will be true for times when $\varepsilon(t)$ for the smooth transition is the same as $\varepsilon_2$ for the step function. However, as discussed above $\varepsilon(t)$ for the smooth transition at times $t > t_0 + \Delta t/2$ will be approximately $\varepsilon_2$. Now, regarding $\tau(t)^{step}$ and $\tau(t)^{smooth}$ from Eq. 9, they only differ by an additional term introduced by the smooth transition, i.e., $-\frac{\varepsilon_2 - \varepsilon_1}{\varepsilon_1} \left[ \frac{1}{\gamma} \ln \left( \frac{\varepsilon_2 + \varepsilon_1 e^{-\gamma(t-t_0)}}{\varepsilon_2 e^{-\gamma t_0} + \varepsilon_1} \right) \right]$, which is time dependent, as expected. From Eq. 8a, the phase of the modal amplitude is $\omega'_1 \tau(t)^{step}$ and $\omega'_1 \tau(t)^{smooth}$, for each function of $n(t)$. This means that there will be specific $\Delta t$ values where the phase for the smooth transition will be approximately the same as that of the time interface; i.e., the modal amplitudes as a function of time will overlap. Considering the full expression for $\omega'_1 \tau(t)^{smooth}$, this occurs when $\frac{\omega'_1}{Z\varepsilon_2} \frac{\varepsilon_2 - \varepsilon_1}{\varepsilon_1} \left[ \frac{1}{\gamma} \ln \left( \frac{\varepsilon_2 + \varepsilon_1 e^{-\gamma(t-t_0)}}{\varepsilon_2 e^{-\gamma t_0} + \varepsilon_1} \right) \right] = 2\pi j$, where $j = 1,2,3 ...$ This yields the following phase condition for $t > t_0 + \Delta t/2$, simplified using the previous considerations $\frac{\varepsilon_1}{\varepsilon_2} \gg e^{-\gamma t_0}$ or $\frac{\varepsilon_1}{\varepsilon_2}, \frac{\varepsilon_2}{\varepsilon_1} \gg e^{-\zeta/2}$ and $t_0 > \Delta t/2$, to mimic the time interface by the smooth transition:

$$\gamma = \frac{\omega'_1}{2\pi j} \frac{\varepsilon_2 - \varepsilon_1}{\varepsilon_1 \varepsilon_2 Z} \ln\left(\frac{\varepsilon_2}{\varepsilon_1}\right), \quad j = 1,2,3, ... \quad (10)$$

This means that if the duration of the smooth transition of $n(t)$ satisfies Eq. 10, the temporal part of electromagnetic field in the cavity produced by the smooth transition of $n(t)$, will approximately match that produced by the time interface (step change of $n(t)$). The implications of Eq. 10 are clearly seen in Fig. 3a,d where the values of $\Delta t/T_1$ (with $\Delta t = 20/\gamma$) corresponding to $j = 1,2,3 ...$ are those where the modal amplitudes ($e(t)^{step} - e(t)^{smooth}$) and the electric field ($E_y(t)^{step} - E_y(t)^{smooth}$) are approximately zero. For completeness, the calculated values of $e(t)^{step} - e(t)^{smooth}$ for $j = 1,2 ... 5$ as a function of $t/T_1$ are shown in Fig. 3b as solid lines. These results demonstrate that after the smooth transition of $n(t)$ is introduced, the modal amplitudes are approximately the same as that of the time interface (step function of $n(t)$). Similarly, we extracted $E_y(t)^{step} - E_y(t)^{smooth}$ as a function of time for the zeroes (or approximately zero values) shown in Fig. 3d and the results are shown in Fig. 3e. Note that the numerical solutions give us approximately but not exactly zero values due to i) the approximate value of $\Delta t/T_1$ for the smooth transitions and ii) the non-zero rising time for the time interface used in the simulations (see methods section for full details). However, it is clear how there is an agreement with the theoretical calculations from Fig. 3b where the smooth transition mimics a time interface. For completeness, the mode amplitudes and electric field (for both the step and smooth transitions) are shown in Fig. 3c,f along with the corresponding functions of $n(t)$ (Fig. 3c), showing how the results (solid lines) overlap with those obtained with a time interface (dashed line). Also, an animation showing the numerical results of $E_y(t)^{\text{step}}$ and $E_y(t)^{\text{smooth}}$ for the case $j = 2$ can be found as a supplementary video.

The results from Fig. 2-3 have shown the case when $n(t)$ is changed from a smaller to a larger value. As a final study, we can also consider the scenario where $n(t)$ is modified from a larger to a smaller value. Most of the parameters for this case are the same as those used in Fig. 2-3 but now with $\varepsilon_1 = 12$, and $\varepsilon_2 = 2$, with the same $Z = 0.85$. Moreover, the eigenfrequency of the cavity is initially $f_1 = 0.25$ GHz, considering the value of $\varepsilon_1 = 12$ for times $t < t_0$ taking the time interface as reference. Note that this is different than the one used in Fig. 2-3 due to the fact that we are keeping the same geometrical dimension of the cavity so the eigenfrequency for times $t < t_0$ will change as now $\varepsilon_1 = 12$. Finally, $T_1 = 4.0$ ns and $t_0/T_1 = 13.25$. As before, we use $t_0 = 53$ ns as the time when the time interface is applied. With this configuration, the theoretical results of the modal amplitudes for the smooth transition of $n(t)$, $e(t)^{\text{smooth}}$, as a function of $\Delta t/T_1$ and time $t/T_1$ are shown in Fig. 4a along with the values of $e(t)^{\text{step}} - e(t)^{\text{smooth}}$ in Fig. 4c (see supplementary materials for the numerical simulations). As observed, there are also values of $\Delta t/T_1$ where the smooth transition of $n(t)$ emulates a time interface. To better observe this, the theoretical and numerical simulations results that satisfy Eq. 10 for $j = 1,2...5$ are shown in Fig. 4b,d, respectively. These results demonstrate that after the time interface is applied, the frequency of the signal inside the cavity is changed to a larger value (Fig. 4a,b), as expected, and $e(t)^{\text{step}} - e(t)^{\text{smooth}} \approx 0$ for $t > t_0$ (Fig. 4c,d). As a final note, the results from Fig. 2-4 correspond to the case when specific values of $\varepsilon_1$ and $\varepsilon_2$ are used. As one would expect, the values of $\Delta t/T_1$ that fulfil Eq. 10 will depend on these parameters. Another example with different values of $\varepsilon_1$ and $\varepsilon_2$ is presented in the supplementary materials, for completeness.

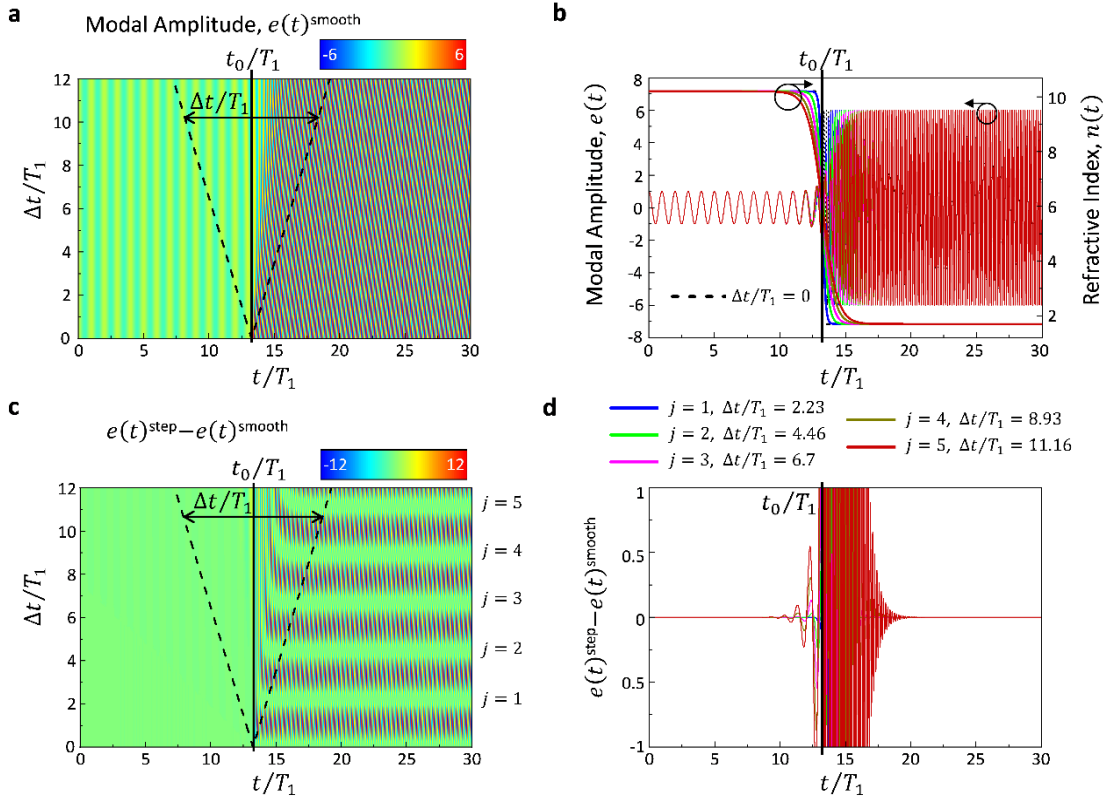

**Fig. 4 Time-dependent fields inside the cavity using smooth transitions and time interfaces via $n(t)$: from larger to smaller values.** Theoretical results of the electric modal amplitude (**a**) as a function of falling time of the smooth transition and time using $\varepsilon_1 = 12$, $\varepsilon_2 = 2$ and time-independent $Z = 0.85$, as in Fig. 2-3. (**c**) Theoretical results of the difference between the electric modal amplitudes produced by step and smooth $n(t)$. (**b**) Electric modal amplitudes related to the time interface, dotted line, and smooth transitions for several values of transition times that fulfil Eq. 10 with the related dependencies of the refractive index, solid lines. (**d**) Theoretical results for the values of $\Delta t/T_1$ from (**c**) where $e^{step} - e^{smooth} \approx 0$. The solid vertical and dashed diagonal lines have the same meaning as those in Figs. 2-3.

### Accuracy implications

As mentioned in the previous section, the parameter $\zeta$ for the smooth transition of $n(t)$ in Eq. 2b, can be used to control how accurate a smooth transition mimics the response of a time interface. For completeness, we provide in Fig. 5 examples of how close the theoretical solution of the modal amplitude of the smooth transition is with respect to that of the time interface (step function). For this, we first theoretically calculated $|e(t)^{step} - e(t)^{smooth}|$ at the time $t = t_0 + \Delta t/2$. Here, differently than the results from Fig. 2-4, it is the parameter $\zeta$ that is changed (ranging from 1 up to 80) while $\gamma$ stays the same; i.e., the same smooth transition is used for all values of $\zeta$. Due to this, $\Delta t$ is recalculated for each value of $\zeta$ as $\Delta t = \zeta/\gamma$. The result of this is that now $\Delta t$ does not provide information about the rising/falling time of the smooth function of $n(t)$ but it gives information of how close $\varepsilon(t)$ for the smooth transition is to $\varepsilon_2$ from the time interface case. In order words, when $\gamma$ is constant, larger values of $\zeta$ will increase $\Delta t$ and, hence, as the observation time will increase ($t =$ $t_0 + \Delta t/2$ ) $\varepsilon(t)$ for the smooth transition will become a better approximation of $\varepsilon_2$ at that time. With this configuration, the results from Fig. 5a and Fig. 5b correspond to the cases when ($\varepsilon_1 = 2, \varepsilon_2 = 12$) and ($\varepsilon_1 = 12, \varepsilon_2 = 2$), respectively, using the first three values of $\gamma$ (Eq. 10) that make both solutions (step and smooth chases) similar; i.e., $j = 1,2,3$. From Eq. 10 $\gamma = 2.24 \times 10^9$ sec$^{-1}$, $\gamma = 1.12 \times 10^9$ sec$^{-1}$ and $\gamma = 7.47 \times 10^8$ sec$^{-1}$ for $j = 1,2,3$, respectively. Meaning that each plot shown in Fig. 5a,b correspond to different smooth transitions that fulfil Eq. 10, while $\gamma$ is kept same for each case when $\zeta$ is modified. As observed in Fig. 5a,b, the difference between the results of the modal amplitudes gets smaller as $\zeta$ increases. This is an expected results as $\zeta = \infty$ will make Eq. 1b converge exactly into Eq. 1a (as explained in the previous section). For the value of $\zeta = 20$ used in the results from Fig. 2-4, the difference between both modal amplitudes is in the order of $\approx 1.8 \times 10^{-5}$ and $\approx 3.1 \times 10^{-4}$ for the cases shown in Fig. 5a,b, respectively, at the time $t = t_0 + \Delta t/2$. However, note that while the difference between modal amplitudes reduces as $\zeta$, this difference starts to saturate for $\zeta > 60$ and $\zeta > 70$ from Fig. 5a,b, which can be due to the computational limitations on the decimal places when calculating the results. Nevertheless, one would expect the difference to asymptotically reach zero for $\zeta = \infty$.

As final study, $t = t_0 + \Delta t/2$ is the time at which the results from Fig. 2-4 and Fig. 5a,b have been obtained. This time has enabled us to compare the potential of smooth transitions to mimic a time interface. As one would expect, if this time is chosen such that $t = t_0 + \alpha$ with $\alpha > \Delta t/2$ and fixed $\gamma$, the results of the modal amplitudes using both functions of $n(t)$ will get much closer as we are allowing more time for the smooth transition to converge in to $\varepsilon_2$. This can be observed in Fig. 5c,d where the influence of $\zeta$ on the accuracy of the emulation when the observation time is changed from $t = t_0 + \Delta t/2$ (as was presented in Fig. 5a,b) to $t = t_0 + 2\Delta t$ for the configurations of $n(t)$ shown in Fig. 2-3 and Fig. 4 is represented in Fig. 5c and Fig. 5d respectively. Here, we consider $j = 1$. From Fig. 5c,d, it can be seen how the modal amplitude for the smooth transitions of $n(t)$ get much closer to the results of the time interface, i.e. the accuracy of the emulation can be further increased by changing $\alpha$. These results demonstrate that, even when the smooth transition does not perfectly match the results of the time interface, it is possible to approximately emulate time interfaces using adiabatic changes of $n(t)$ by carefully choosing the rise/fall time of the transition as well as the observation time. These results may open further experimental routes to exploit time interfaces being emulated by adiabatic modulations of $n(t)$.

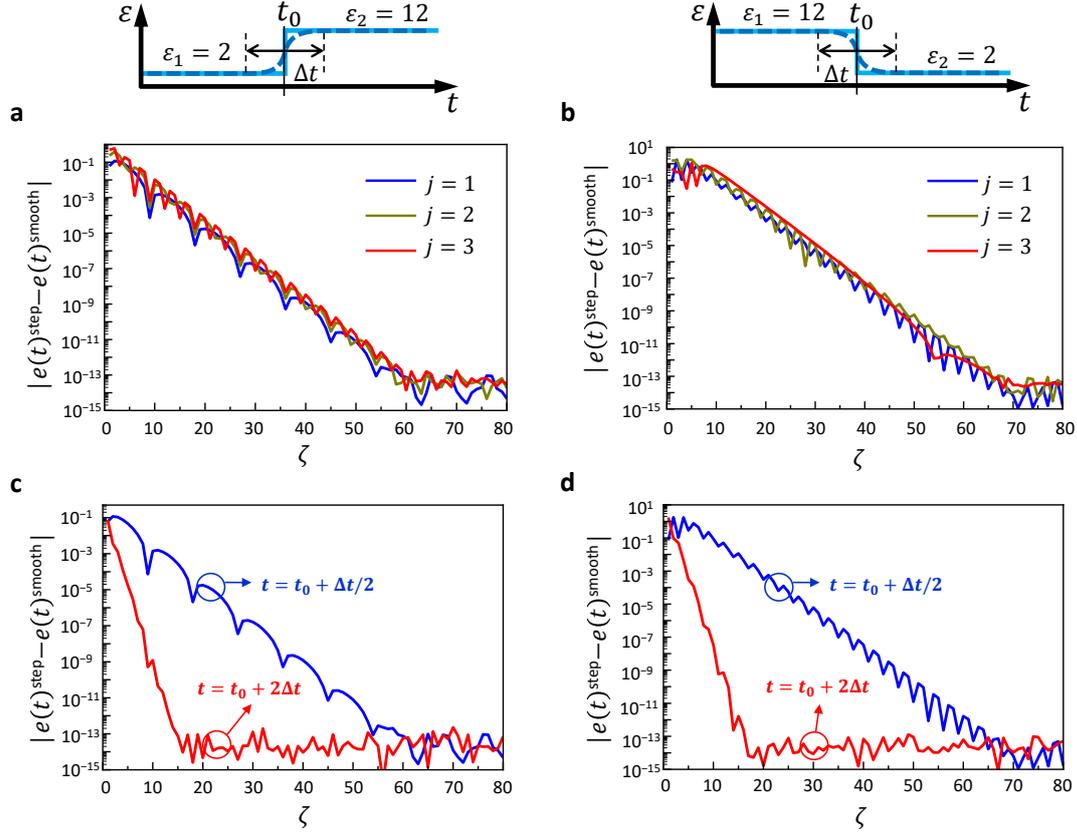

**Fig. 5 Estimation of accuracy of the emulation of time interfaces via the smooth transitions. (a,b)** Theoretical results of the absolute value of the difference between the electric modal amplitudes produced by the step and smooth functions when $n(t)$ is increased or reduced in time, respectively. These results are calculated at the time $t = t_0 + \Delta t/2$ for the condition $|e^{step} - e^{smooth}| \approx 0$ ($j = 1,2,3$). **(c,d)** Values of $|e(t)^{step} - e(t)^{smooth}|$ produced by the step and smooth transitions for the same $\varepsilon_1$ and $\varepsilon_2$ and $j = 1$ as in panels **(a,b)** respectively, as function of the parameter $\zeta$. These results are calculated using $t = t_0 + \Delta t/2$ (blue) and $t = t_0 + 2\Delta t$ (red).

**Conclusions**

In conclusion, we have discussed the potential of adiabatic changes of $n(t)$ to mimic the response of time interfaces. Full theoretical derivations have been carried out demonstrating the conditions of rising/falling time of the smooth transition that can enable this. Our findings may open further opportunities to explore in the realm of spacetime media. By enabling the emulation of time interfaces via adiabatic modulations of $n(t)$, the results here presented may serve as the basis to explore further experimental demonstrations by relaxing the need of ultra-fast modulation of the properties of the material where a signal travels.

**Methods**

**Amplitude matching condition.**

Consider the amplitude $A(t)^{\text{smooth}}$, obtained for $t > t_0$ (Eq. 9b) evaluated at longer times $t > t_0 + \frac{\Delta t}{2}$. Taking into account the relation between the control parameter $\gamma$ of the smooth transition and the estimation of the transition time $\Delta t = \zeta/\gamma$, we get:

$$A(t)^{\text{smooth}}\Big|_{t=t_0+\frac{\Delta t}{2}} = \frac{(\varepsilon_2+\varepsilon_1 e^{\gamma t_0})(1+e^{-\gamma \Delta t/2})}{(1+e^{\gamma t_0})(\varepsilon_2+\varepsilon_1 e^{-\gamma \Delta t/2})} = \frac{(\varepsilon_2+\varepsilon_1 e^{\gamma t_0})(1+e^{-\zeta/2})}{(1+e^{\gamma t_0})(\varepsilon_2+\varepsilon_1 e^{-\zeta/2})} \quad (11)$$

At $t = t_0 + \frac{\Delta t}{2}$, Eq. 11 becomes $A(t)^{\text{smooth}}\Big|_{t=t_0+\frac{\Delta t}{2}} = \frac{(\varepsilon_2+\varepsilon_1 e^{\gamma t_0})}{\varepsilon_2(1+e^{\gamma t_0})}$. This means that the conditions that enables $A(t)^{\text{smooth}}\Big|_{t=t_0+\frac{\Delta t}{2}} = A(t)^{\text{step}}\Big|_{t>t_0} = \frac{\varepsilon_1}{\varepsilon_2}$ to hold requires $e^{\gamma t_0} \gg 1$ and $\varepsilon_1 e^{\gamma t_0} \gg \varepsilon_2$ for the denominator and numerator, respectively, of $A(t)^{\text{smooth}}\Big|_{t=t_0+\frac{\Delta t}{2}}$. The first condition can be reduced to $\gamma t_0 \gg 0$ which can be rewritten as $\frac{\zeta}{\Delta t} t_0 \gg 0$. The second condition corresponds to the relation $\frac{\varepsilon_1}{\varepsilon_2} \gg e^{-\gamma t_0}$ shown in the main text.

**Numerical simulations.**

The numerical simulations were carried out using the time-domain solver of the software COMSOL Multiphysics® with a similar setup as in [18]. To excite the cavity, a total of 17 vertically polarized dipoles were placed $0.05\lambda_1$ away from the left PEC boundary of the cavity separated by a distance of $0.05\lambda_1$ along the $y$ axis. Here $\lambda_1$ is the wavelength of the signal inside the cavity when $\boldsymbol{\varepsilon(t) = \varepsilon_1}$. To appreciate the influence of the time interface and the smooth transition of $\boldsymbol{n(t)}$, the dipoles are first switched off and then, after some time, $\boldsymbol{n(t)}$ of the medium within the cavity is changed. For instance, for the results from Fig. 2, the dipoles are switched off at $\boldsymbol{t \approx 12T_1}$ and the time interface is applied at $\boldsymbol{t \approx 79.1T_1}$. A triangular mesh with maximum and minimum sizes as in [18] was implemented. For the time interface, $\boldsymbol{\varepsilon(t), \mu(t)}$ were implemented using analytical functions having a rise/fall time of $1.5\times10^{-3}\boldsymbol{T_1}$ with a smoothing of two continuous derivatives. The smooth transitions were implemented using the analytical expression described in Eq. 1b.

**Acknowledgements**

V.P.-P. would like to thank the support of the Leverhulme Trust under the Leverhulme Trust Research Project Grant scheme (No. RPG-2020-316 and RPG-2023-024). M.A. acknowledges the British Academy and CARA for support through the Researcher at Risk Fellowship programme. For the purpose of Open Access, the authors have applied a CC BY public copyright license to any Author Accepted Manuscript (AAM) version arising from this submission.

**Conflicts of interests**

The authors declare no conflicts of interests.